\documentclass[english,conference]{IEEEtran}
\usepackage[latin9]{inputenc}
\usepackage{color}

\usepackage{amsthm}
\usepackage{amsmath}
\usepackage{amssymb}
\usepackage{graphicx}

\newtheorem{thm}{Theorem}
\newtheorem{cor}{Corollary}
\newtheorem{lem}{Lemma}
\newtheorem{defn}{Definition}
\newtheorem{prop}{Proposition}

\usepackage[english]{babel}
\usepackage{enumerate}\usepackage{verbatim}\usepackage{xcolor}\usepackage{bbm}
\usepackage{subcaption}
\usepackage{tikz}
\usetikzlibrary{arrows,positioning,fit,backgrounds}
\usetikzlibrary{calc}

\usepackage{color}   
\usepackage{hyperref}
\hypersetup{
    colorlinks=true, 
    linktoc=all,     
    linkcolor=blue,  
}

\newcommand{\Ebb}{\mathbb{E}}
\newcommand{\Pbb}{\mathbb{P}}
\newcommand{\Acal}{\mathcal{A}}
\newcommand{\Ccal}{\mathcal{C}}

\newcommand{\Scal}{\mathcal{S}}
\newcommand{\Tcal}{\mathcal{T}}
\newcommand{\Zcal}{\mathcal{Z}}
\newcommand{\zsf}{\mathsf{z}}

\tikzstyle{arw}=[->,>=latex]
\tikzstyle{node}=[draw,rectangle,rounded corners, minimum width=1cm,minimum height =.75 cm]

\usepackage{adjustbox}

\begin{document}

\title{Secrecy Communication with Security Rate Measure}

\author{Lei Yu, Houqiang Li, and Weiping Li, \emph{Fellow, IEEE}\\
 University of Science and Technology of China\\
 Email: yulei@mail.ustc.edu.cn, \{lihq,wpli\}@ustc.edu.cn}

\maketitle
\begin{abstract}
We introduce a new measure on secrecy, which is established based
on rate-distortion theory. It is named \emph{security rate}, which
is the minimum (infimum) of the additional rate needed to reconstruct
the source within target distortion level with any positive probability
for wiretapper. It denotes the minimum distance in information metric
(bits) from what wiretapper has received to any decrypted reconstruction
(where decryption is defined as reconstruction within target distortion
level with some positive possibility). By source coding theorem, it
is equivalent to a distortion-based equivocation $\mathop{\min}\limits _{p\left(v^{n}|s^{n},m\right):Ed\left(S^{n},V^{n}\right)\le D_{E}}\frac{1}{n}I\left(S^{n};V^{n}|M\right)$
which can be seen as a direct extension of equivocation $\frac{1}{n}H\left(S^{n}|M\right)$
to lossy decryption case, given distortion level $D_{E}$ and the
received (encrypted) message $M$ of wiretapper. In this paper, we
study it in Shannon cipher system with lossless communication, where
a source is transmitted from sender to legitimate receiver secretly
and losslessly, and also eavesdropped by a wiretapper. We characterize
the admissible region of secret key rate, coding rate of the source,
wiretapper distortion, and security rate (distortion-based equivocation).
Since the security rate equals the distortion-based equivocation,
and the equivocation is a special case of the distortion-based equivocation
(with Hamming distortion measure and $D_{E}=0$), this gives an answer
for the meaning of the maximum equivocation.
\end{abstract}



\section{Introduction}

\begin{figure}
\centering \includegraphics[width=0.4\textwidth,height=0.15\textwidth]{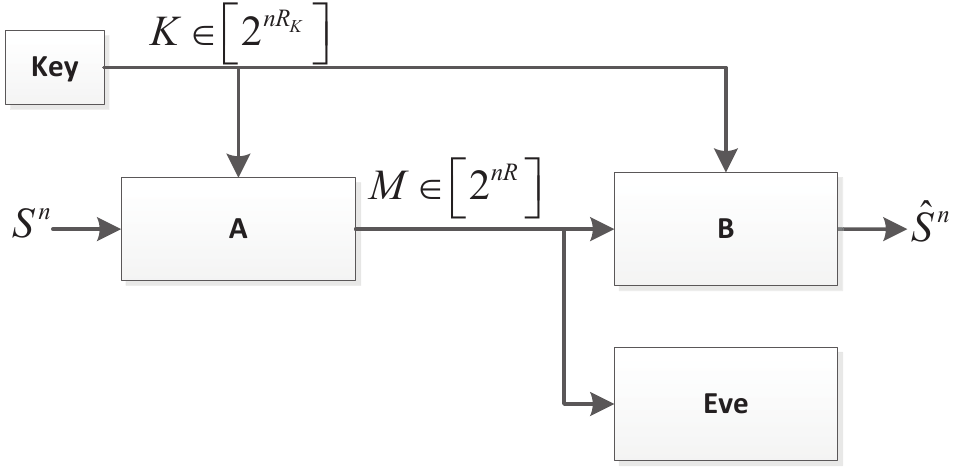}
\protect\protect\protect\protect\caption{\label{fig:Shannon} The Shannon cipher system. }
\end{figure}

The Shannon cipher system depicted in Fig. \ref{fig:Shannon} is first
investigated in \cite{Shannon}, where sender A communicates with
legitimate receiver B secretly by exploiting a secret key that is
shared by them. In \cite{Shannon}, Shannon regarded this system as
perfectly secret if the source and the eavesdropped message are statistically
independent. For lossless communication case, a necessary and sufficient
condition for perfect secrecy is that the number of secret key bits
per source symbol exceeds the entropy of the source. When the amount
of key is insufficient, it must relax the requirement of statistical
independence and introduce new measures of secrecy.

One common way of measuring sub-perfect secrecy is with equivocation,
the conditional entropy $\frac{1}{n}H\left(S^{n}|M\right)$ of the
source given the public message. The use of equivocation as a measure
of secrecy was considered in the original work on the wiretap channel
in \cite{Wyner} and \cite{Csiszar}, and it continues today. From
lossy source coding theorem, the equivocation indeed indicates the
minimum additional rate for wiretapper to reconstruct the source losslessly
on the basis of the received message $M$, when the coding scheme
adopted by A and B is stationary.

For lossy decryption, a distortion-based measure has been proposed
by Yamamoto in \cite{Yamamoto}, which is to measure secrecy by the
distortion that an wiretapper incurs in attempting to reconstruct
the source sequence. However, Schieler et al \cite{Schieler13} show
that this measure is cheap, since negligible rates of secret key can
force minimum distortion achieved by wiretapper to the one reconstructed
without any information about the source; meanwhile it is also fragile,
since wiretapper can reconstruct the transmitted message completely
if it has a little knowledge of the source. To strengthen measure
of secrecy, another distortion-based approach \cite{Schieler14},
\cite{Schieler14-2} is to design schemes around the assumption that
the wiretapper has ability to conduct list decoding with fixed list
size. The induced distortion is the minimum distortion over the entire
list. This exponent of list size $R_{L}$ is an important indicator
to secrecy performance. From aspect of uncertainty, $R_{L}$ indeed
indicates the minimum additional rate needed to reconstruct the source
within target distortion level with ``zero-delay'' decryption constraint,
where ``zero-delay'' follows from that only single-block codes are
allowed for wiretapper, i.e., the blocklength adopted by wiretapper
is restricted to be the same to that adopted by A and B (in fact,
without delay constraint, the wiretapper would collect multiple blocks
to produce a supperblock reconstruction, which is called \emph{superblock
coding}). This quantity is somewhat similar to equivocation (i.e.,
$\frac{1}{n}H\left(S^{n}|M\right)$), which is also to measure the
uncertainty of the wiretapper. However the equivocation is the minimum
additional rate to decrypt for henchman and wiretapper, when wiretapper
could apply arbitrary length supercode; while for $R_{L}$, only single-block
codes are allowed for wiretapper. In addition, from aspect of complexity,
$R_{L}$ indicates the exponent of admissible maximum complexity (exponent
of admissible maximum list size over all blocks) needed to reconstruct
the source within target distortion level.

In this paper, we study the Henchman problem with superblock coding,
and define security rate as the minimum (infimum) of the additional
rate needed to reconstruct the source within target distortion level
with any positive probability for wiretapper. We aim at characterizing
the optimal tradeoff among secret key rate, coding rate of the source,
wiretapper distortion $D_{E}$, and security rate. An important difference
from \cite{Schieler14} is that we do not constrain the blocklength
adopted by the Henchman and wiretapper; while in \cite{Schieler14},
the blocklength adopted by the Henchman and wiretapper is restricted
to be the same to that adopted by A and B. It is worth noting that
the proof in \cite{Schieler14} relies on a likelihood encoder by
using the soft covering lemma \cite{Cuff13}, which is invalid to
solve our problem. This is because if consider the superblock case
and let $l\rightarrow\infty$ before let $n\rightarrow\infty$, then
the soft covering lemma does not hold, where $l$ denotes the number
of subblocks in each supperblock, and $n$ denotes the number of symbols
in each subblock (which is also the blocklength of the code adopted
by A and B). Actually, we find a more general approach to solve our
problem which relies on jointly typical coding. Moreover, our approach
is available to prove the results of \cite{Schieler14}. In addition,
we also define distortion-based equivocation as $\mathop{\min}\limits _{p\left(v^{n}|s^{n},m\right):Ed\left(S^{n},V^{n}\right)\le D_{E}}\frac{1}{n}I\left(S^{n};V^{n}|M\right)$,
which can be seen as an extension of equivocation $\frac{1}{n}H\left(S^{n}|M\right)$
to lossy decryption case. By source coding theorem, in our problem
the security rate is exactly equal to the distortion-based equivocation.
Hence utilizing the result on security rate, we can easily characterize
the admissible region of secret key rate, coding rate of the source,
wiretapper distortion $D_{E}$ and distortion-based equivocation by
replacing security rate with distortion-based equivocation. It means
that maximizing equivocation is indeed equivalent to maximizing security
rate of lossless decryption. We also extend our results on lossless
communication case to lossy communication case.

\begin{figure}
\centering \includegraphics[width=0.4\textwidth,height=0.15\textwidth]{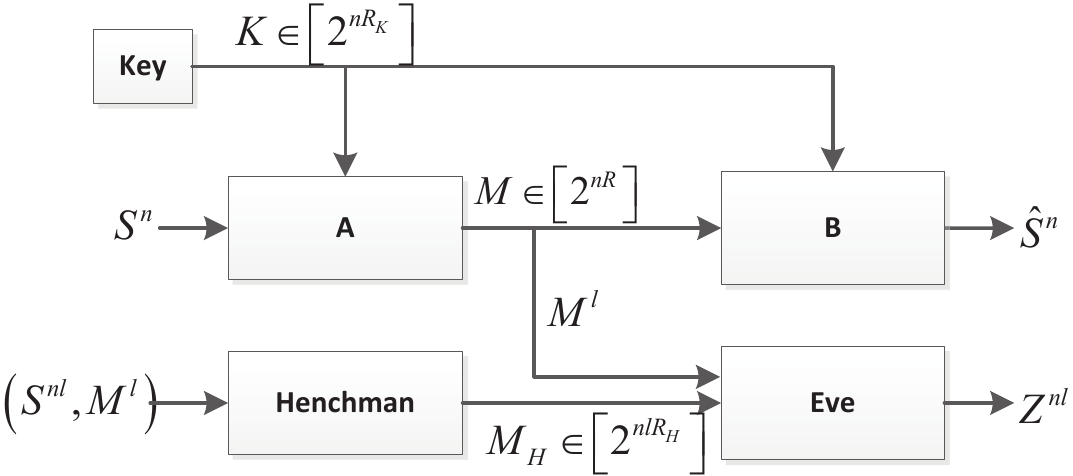}
\protect\protect\protect\protect\caption{\label{fig:henchman}The henchman problem with superblock coding.\emph{
}A henchman who has access to the $l$ blocks of source sequence, $S^{nl}$,
and the $l$ blocks of public message, $M^{l}$, codes them with rate $R_H$ to help wiretapper to produce a reconstruction.
The wiretapper reconstructs a sequence $Z^{nl}$ based on the public message $M^{l}$ and the information
$M_{H}$ from the henchman. Here $M^{l}=\left(M_{1},M_{2},\cdots,M_{l}\right)$
denotes a sequence of the messages $M$. }
\end{figure}

\section{\label{sec:Problem-Formulation}Problem Formulation and Preliminaries}

Consider the secrecy communication system shown in Fig. \ref{fig:Shannon}.
The sender A observes a source sequence $S^{n}$ that is i.i.d. according
to a distribution $P_{S}$ . Sender A and legitimate receiver B share
a secret key $K$ that is uniformly distributed in $\left[2^{nR_{K}}\right]$
\footnote{In this paper, the set ${1,...,m}$ is sometimes denoted by $[m]$.%
} and independent of $S^{n}$. Sender A encodes the source using the
secret key and then sends the coded message $M$ to legitimate receiver
B over a noiseless channel at rate $R$, which is also wiretapped
by a wiretapper Eve. To make our problem more clear, we introduce
the Henchman problem with supperblock coding shown in Fig. \ref{fig:henchman}
(the single block coding version is investigated in \cite{Schieler14}).

\subsection*{\textit{Henchman problem with supperblock coding}}
\begin{defn}
The encoder and decoder of a $\left(n,R,R_{K}\right)$ block code
are defined by the following two mappings:
\begin{equation}
\begin{array}{l}
{Encoder\; f:{\mathcal{S}}^{n}\times\left[2^{nR_{K}}\right]\to\left[2^{nR}\right]}\\
{Decoder\; g:\left[2^{nR}\right]\times\left[2^{nR_{K}}\right]\to\hat{{\mathcal{S}}}^{n}}
\end{array}\label{1)-1-1}
\end{equation}
\end{defn}
The wiretapper Eve overhears the encrypted message $M$ perfectly.
A and B want to communicate within certain distortion level by using the secret key and the noiseless channel, while ensuring that the wiretapper
suffers distortion above a certain threshold with high probability.
We say a source sequence $s^{n}$ is decrypted by wiretapper if and
only if the wiretapper produces a reconstruction of $s^{n}$, $z^{n}$,
such that
\begin{equation}
d\left(s^{n},z^{n}\right)\le D_{E}\label{4)-2-1}
\end{equation}
When $D_{E}$ is small enough, and A and B adopt an appropriate coding
scheme, then the wiretapper cannot decrypt the source only by $M$.
Hence, we can measure security by the minimum additional bit rate
needed for wiretapper to decrypt the source. In this case, it is equivalent
to that there is a rate-limited helper (i.e., a henchman) who can
access all information about source (source $S^{n}$ and the encrypted
message $M$). As depicted in Fig. \ref{fig:henchman}, by applying
an $l$- length of superblock code (each subblock with $n$ symbols),
the wiretapper receives $nlR_{H}$ bits of side information from a
henchman who has access to the source sequence $S^{nl}$ and the public
message $M^{l}$. Since the wiretapper and henchman cooperate, this
means that the wiretapper effectively receives the best possible $nlR_{H}$
bits of side information about the pair $\left(S^{nl},M^{l}\right)$
to assist in producing a reconstruction sequence $Z^{nl}$.
\begin{defn}
The $\left(l,R_{H}\right)$ Henchman code (Hcode) of a $\left(n,R,R_{K}\right)$
block code is a superblock defined by the following two mappings:
\begin{equation}
\begin{array}{l}
{Encoder\; f_{H}:{\mathcal{S}}^{nl}\times\left[2^{nlR}\right]\to\left[2^{nlR_{H}}\right]}\\
{Decoder\; g_{H}:\left[2^{nlR_{H}}\right]\times\left[2^{nlR}\right]\to\mathcal{Z}^{nl}}
\end{array}\label{10)}
\end{equation}
The encoder and decoder of Hcode can be stochastic.
\end{defn}
For any $\left(n,R,R_{K}\right)$ block code, we define security rate
to measure its secrecy.
\begin{defn}
\label{def:SecRate}For an $\left(n,R,R_{K}\right)$ block code adopted
by A and B, and a given decryption distortion level $D_{E}$, Henchman
rate $R_{H}$ of the $\left(n,R,R_{K}\right)$ block code is said
to be secure if for $\forall\epsilon>0$, the following inequality
holds. In this case, $R_{H}$ is said to be a secure Henchman rate
or security rate.

\begin{equation}
\mathop{\limsup}\limits _{l\to\infty}\mathop{\max_{\begin{array}{c}
\left(l,R'_{H}\right)Hcodes\\
R'_{H}\leq R_{H}-\epsilon
\end{array}}}\mathbb{P}\left[d\left(S^{nl},Z^{nl}\right)\le D_{E}\right]\leq\epsilon\label{eq:SecRate}
\end{equation}

\end{defn}
From Definition \ref{def:SecRate}, we have that if the wiretapper
receives any additional message with rate $\leq$ security rate, then
it cannot reconstruct the source within target distortion level with
probability$\geq\epsilon$ for any $\epsilon>0$. Security rate also
denotes the infimum of the additional rate needed to reconstruct the
source within target distortion level with probability$\geq\epsilon$
for any $\epsilon>0$ for wiretapper.

To keep the transmission as secret as possible from Eve, A and B should
maximize security rate or minimize decryption probability in \eqref{eq:SecRate}.
\begin{defn}
The tuple $\left(R,R_{K},R_{H},D_{E}\right)$
is achievable for distortion measure $d$ if for $\forall\epsilon>0$,
as well as sufficiently large $n$, there exists an $\left(n,R',R'_{K}\right)$
code satisfying
\begin{eqnarray}
R'_{K} & \le & R_{K}+\epsilon\label{101)}\\
R' & \le & R+\epsilon\\
\mathbb{P}\left[S^{n}\neq\hat{S}^{n}\right] & \le & \epsilon\\
R_{H}\left(D_{E}\right) & \geq & R_{H}-\epsilon\label{eq:SecrRateCons}
\end{eqnarray}
Inequality \eqref{eq:SecrRateCons} is equivalent to \eqref{eq:SecRate}.
\end{defn}

\begin{defn}
\label{def:MaxSR} Given $R$, $R_{K}$,
and $D_{E}$, maximum security rate $R_{H}^{*}\left(R,R_{K},D_{E}\right)$
is defined as the supremum of all $R_{H}$ such that $\left(R,R_{K},R_{H},D_{E}\right)$
is achievable.
\end{defn}
The wiretapper and henchman jointly design a code consisting of an
encoder $m_{H}=f_{H}\left(s^{nl},m^{l}\right)$ and a decoder $z^{nl}=g_{H}\left(m_{H},m^{l}\right)$,
subject to the constraint $\left|M_{H}\right|\le2^{nlR_{H}}$. We assume that
the henchman and the wiretapper are both aware of the scheme that Nodes A and B employ, but neither aware of the secret key.
That is to say the $f_{H}\left(s^{nl},m^{l}\right),g_{H}\left(m_{H},m^{l}\right)$
may be designed based on the $\left(n,R,R_{K}\right)$ block code.

Although we assume the excess distortion constraint constraint of
distortion for wiretapper here, from the proofs of our results, our
results still hold if apply expected distortion constraint for wiretapper
\footnote{Assume the distortion function is upper bounded.%
}. 

In addition, it is worth noting that based on the definitions above,
maximum security rate is equivalent to the following definition.
\begin{defn}
\textit{\label{def:DistBasedEqui}} For a $\left(n,R,R_{K}\right)$
block code adopted by A and B, and for a given decryption distortion
level $D_{E}$, distortion-based equivocation is defined as:

\begin{equation}
{R_{DE}\left(D_{E}\right)=\mathop{\min}\limits _{\begin{array}{c}
p\left(v^{n}|s^{n},m\right):\\
Ed\left(S^{n},V^{n}\right)\le D_{E}
\end{array}}\frac{1}{n}I\left(S^{n};V^{n}|M\right)}\label{eq:DistBasedEqui}
\end{equation}

\end{defn}
The distortion-based equivocation is exactly a direct extension of
equivocation to lossy decryption case. For maximum security rate in
Definition \ref{def:MaxSR} we restrict the code adopted by A and
B to be stationary (independent of time), hence the outputs $M^{l}$
must be a stationary process. Then by source coding theorem, $R_{DE}\left(D_{E}\right)$
is the minimum additional rate to decrypt source with any positive
probability for henchman and wiretapper. Therefore, maximum $R_{DE}\left(D_{E}\right)$
is exactly equal to $R_{H}^{*}\left(R,R_{K},D_{E}\right)$. This implies
that maximum equivocation is equal to the maximum security rate of
lossless decryption.

\section{\label{sec:Main-Results}Main Results}

\begin{figure}
\centering \includegraphics[width=0.5\textwidth,height=0.3\textwidth]{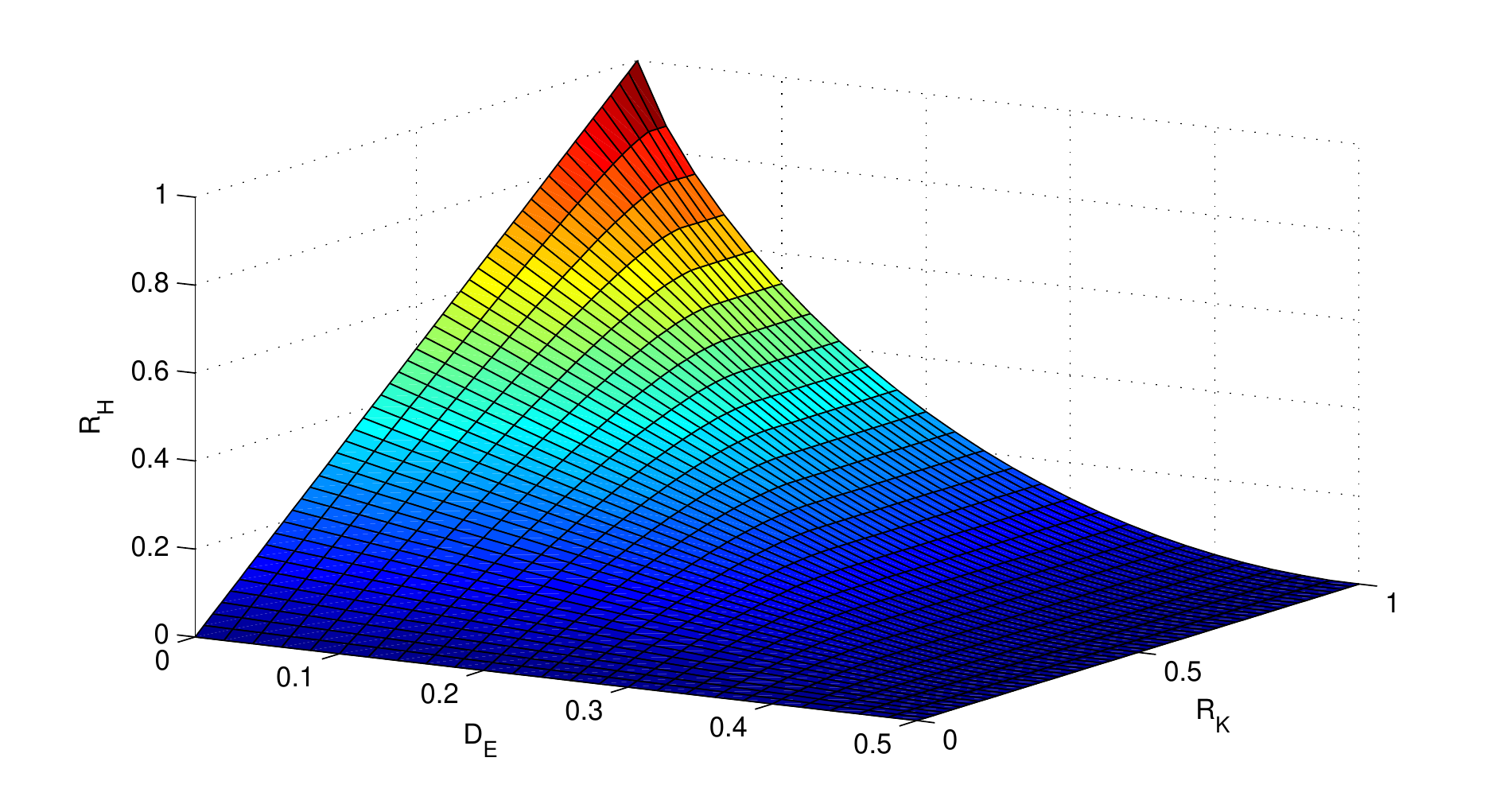}
\protect\protect\protect\protect\caption{\label{fig:Thm1}The region in Theorem \ref{thm:Thm1} for source
$P_{S}\sim$Bern(1/2) and distortion measure $d(s,z)=1\left\{ s\protect\neq z\right\} $. }
\end{figure}

The following theorem characterizes the admissible region for lossless
communication among the communication rate, secret key rate, wiretapper
distortion, and security rate.
\begin{thm}
\label{thm:Thm1}Given a source distribution $P_{S}$ and distortion
measure $d$, the tuple $\left(R,R_{K},R_{H},D_{E}\right)$ is achievable
if and only if it holds that

\begin{align}
\begin{array}{l}
R\end{array} & \ge H\left(S\right),\label{eq:Thm1-1}\\
R_{H} & \le\Gamma\left(R_{K},D_{E}\right)\label{eq:Thm1-2}
\end{align}
where
\begin{equation}
\Gamma\left(R_{K},D_{E}\right)\triangleq\mathop{\min}\limits _{\begin{array}{l}
(\lambda,D):{0\le\lambda\le1,}\\
{\lambda D\le D_{E}}
\end{array}}{(1-\lambda)R_{K}+\lambda R(D)}\label{eq:Gamma}
\end{equation}
$R\left(D\right)$ denotes rate-distortion function of source $S$
with distortion measure $d\left(s,\hat{s}\right)$, i.e.,
\begin{equation}
\begin{array}{l}
{R\left(D\right)=\mathop{\min}\limits _{p\left(v|s\right):Ed\left(S,V\right)\le D}I\left(S;V\right)}\end{array}\label{ZEqnNum861486-1}
\end{equation}

\end{thm}
The function \eqref{eq:Gamma} is exactly the rate-distortion tradeoff
of the timesharing between the points $\left(0,R_{K}\right)$ and
$\left(D,R\left(D\right)\right)$. Note that the region of Theorem
\ref{thm:Thm1} is different from that in Theorem 1 of \cite{Schieler14},
which is the singleblock coding version of Henchman problem. More
precisely, the region of Theorem \ref{thm:Thm1} is the convex hull
of that in Theorem 1 of \cite{Schieler14} in $R_{H}$ and $D_{E}$
dimensions. This is because the timesharing is feasible for superblock
coding, but not feasible for singleblock coding. It also implies that
timesharing is one of the optimal strategies to achieve the optimal
$R_{H}$ and $D_{E}$ tradeoff. Fig. \ref{fig:Thm1} illustrates Theorem~\ref{thm:Thm1}
for a Bern(1/2) source and Hamming distortion; the communication rate
is assumed to satisfy $R\geq H(S)$, and they have no effect on the
$\left(R_{K},R_{H},D_{E}\right)$ tradeoff. In Fig. \ref{fig:Thm1},
for given $R_{K}$ and $R_{H}$, the distortion region below the surface
is not achievable for wiretapper with probability 1; while the distortion
region above the surface is achievable for wiretapper with probability
1.

We can readily establish the following corollaries to Theorem~\ref{thm:Thm1}.
\begin{cor}
\label{thm:Cor1-1}Given a source distribution $P_{S}$ and distortion
measure $d$, for all $\begin{array}{l}
R\end{array}\ge H\left(S\right)$,

\begin{equation}
R_{H}^{*}\left(R,R_{K},D_{E}\right)=\Gamma\left(R_{K},D_{E}\right)\label{eq:Thm1-2-2-1}
\end{equation}

\end{cor}
From the fact maximum $R_{DE}\left(D_{E}\right)$ is equal to $R_{H}^{*}\left(R,R_{K},D_{E}\right)$,
we have the following corollary.
\begin{cor}
\label{cor:DisEqu}Given $R\geq H\left(S\right)$ and $R_{K}$, maximum\textup{
$R_{DE}\left(D_{E}\right)$} equals the $R_{H}^{*}\left(R,R_{K},D_{E}\right)$,
i.e., they satisfy

\begin{align}
 & \mathop{\limsup}\limits _{n\to\infty}\mathop{\max_{\begin{array}{c}
\left(n,R',R'_{K}\right)codes\\
R'\leq R,R'_{K}\leq R_{K}
\end{array}}}R_{DE}\left(D_{E}\right)\nonumber \\
= & R_{H}^{*}\left(R,R_{K},D_{E}\right)\\
= & \Gamma\left(R_{K},D_{E}\right)
\end{align}

\end{cor}
In addition, for lossless reconstruction at wiretapper, observe that
$\mathop{\limsup}\limits _{n\to\infty}\mathop{\max_{\begin{array}{c}
\left(n,R',R'_{K}\right)codes\\
R'\leq R,R'_{K}\leq R_{K}
\end{array}}}\frac{1}{n}H\left(S^{n}|M\right)$ equals $\mathop{\limsup}\limits _{n\to\infty}\mathop{\max_{\begin{array}{c}
\left(n,R',R'_{K}\right)codes\\
R'\leq R,R'_{K}\leq R_{K}
\end{array}}}R_{DE}\left(0\right)$ with Hamming distortion measure, hence by Corollary \ref{cor:DisEqu}
the maximum equivocation is indeed equal to the maximum security rate.
\begin{cor}
\label{cor:Equ}Given $R\geq H\left(S\right)$ and $R_{K}$, maximum
equivocation equals the $R_{H}^{*}\left(R,R_{K},D_{E}=0\right)$,
with Hamming distortion measure and also equals $\Gamma\left(R_{K},D_{E}=0\right)$.
\end{cor}
Note that when setting $d\left(s,\hat{s}\right)=1\left\{ s\ne\hat{s}\right\} $
in the region of Corollary \ref{cor:Equ}, corresponds to requiring
a lossless reconstruction at wiretapper, which was Shannon's original
formulation of the problem in \cite{Shannon}. In this case, we see
that the tuple $\left(R,R_{K},R_{H},D_{E}=0\right)$ is achievable
if and only if
\begin{align}
R & \ge H\left(S\right)\\
\mathop{\limsup}\limits _{n\to\infty}\mathop{\max_{\begin{array}{c}
\left(n,R',R'_{K}\right)codes\\
R'\leq R,R'_{K}\leq R_{K}
\end{array}}}R_{DE}\left(0\right) & =\min\left\{ R_{K},H\left(S\right)\right\}
\end{align}
This implies the Shannon's results \cite{Shannon}, i.e., the perfect
secrecy (equivocation equals $H\left(S\right)$) is achievable if
and only if $R_{K}\ge H\left(S\right)$.

We now prove the achievability and converse parts of Theorems \ref{thm:Thm1}.

\section{Converse of Theorem \ref{thm:Thm1}}

\label{sec:losslessconverse} The constraint $R\geq H(S)$ follows
from the lossless source coding theorem. If $R_{H}\geq R_{K}$, then
the henchman can send the index of possible decryptions of $M$. For
any scheme that Nodes A and B use, Node B can always use the rate-$R_{K}$
key to identify the correct decryption of $M$ (with probability 1),
hence the number of possible decryptions of $M$ is at most $2^{nR_{K}}$.
This means that it is possible for the wiretapper to produce a lossless
reconstruction (with probability 1) with henchman rate $R_{H}$. On
the other hand, if $R_{H}\geq R\left(D_{E}\right)$, then the henchman
and the wiretapper can simply use a point-to-point
rate-distortion code (ignore $M$ altogether) to describe $S^{n}$ within distortion $D_{E}$
(with probability 1), no matter what scheme Nodes A and B use. In
addition, by applying time-sharing, the the henchman and the wiretapper
are able to achieve any distortion-rate pair on or above the convex
hull of $\left(D,\min\left\{ R_{K},R\left(D\right)\right\} \right)$.
Hence, to prevent wiretapper from achieving that, the inequality\eqref{eq:Thm1-2}
holds.

\section{\label{sec:Achievability-of-Theorem}Achievability of Theorem \ref{thm:Thm1}}

To prove the achievability, we only need to prove that if \eqref{eq:Thm1-1}
and \eqref{eq:Thm1-2} hold, then there exists a code (adopted by
A and B) making \eqref{eq:SecRate} hold. It means that the henchman
observes the pair $\left(S^{nl},M^{l}\right)$ and encodes a message
$M_{H}$, and the wiretapper observes $\left(M^{l},M_{H}\right)$
and decodes $Z^{nl}$; their goal is to minimize the distortion $d\left(S^{nl},Z^{nl}\right)$.
This is just the usual rate-distortion setting with side information
$M^{l}$ available at both encoder and decoder. We will prove that
if the Node A and B use the following coding scheme to encode and
decode the source sequence, then \eqref{eq:SecRate} holds.

Generation of codebooks: First, randomly generate a codebook ${\mathbf{\mathcal{C}}}_{S}$
consisting of $2^{nR}$ sequences $S^{n}$ drawn i.i.d. $\sim\prod_{i=1}^{n}P_{S}\left(s_{i}\right)$.
Then divide all the codewords into bins of size $2^{nR_{K}}$. Index
bins by $j_{p}\in\left[2^{n\left(R-R_{K}\right)}\right]$ and denote
them as ${\mathcal{C}}_{S}\left(j_{p}\right)$. Also index codewords
of each bin by $j_{s}\in\left[2^{nR_{K}}\right]$. The codebook as
well as binning and indexing are known to everyone, including the
adversaries (henchman and wiretapper).

Encoding at sender: Sender encode the sequence $s^{n}$ by $j=\left(j_{p},j_{s}\right)$
if there exists an $\left(j_{p},j_{s}\right)$ pair such that $s^{n}=s^{n}\left(j_{p},j_{s}\right)$
in codebook ${\mathcal{C}}_{S}$ and $s^{n}\left(j_{p},j_{s}\right)\in\Tcal_{\delta}^{n}$
(This is equivalent to find an $\left(j_{p},j_{s}\right)$ pair such
that $\left(s^{n},s^{n}\left(j_{p},j_{s}\right)\right)\in\Tcal_{\delta}^{n}\left(S,\hat{S}\right)$,
where test channel $P_{\hat{S}|S}\left(\hat{s}|s\right)=1\left\{ \hat{s}=s\right\} $).
If there is more than one such index, randomly (uniformly) choose
one among them. If there is no such index, randomly choose one index
from $\left[2^{nR}\right]$. Then the index within that bin, $j_{s}$,
is one-time padded with the key sequence $k$. Denote the message
to $m=\left(j_{p},m_{s}\right)$, where $m_{s}$ denotes the resulting
message by one-time pad operation on $j_{s}$ with $k$.

Decoding at legitimate receiver: Using the key $k$, legitimate receiver
achieves $\left(j_{p},j_{s}\right)$ from the received $m$, and then
it reproduces the sequence $s^{n}\left(j_{p},j_{s}\right)$.

By the standard proof on lossless source coding using joint typicality
coding, we can readily establish that if $R>H\left(S\right)$, then
the legitimate receiver can reconstruct $S^{n}$ with high probability.
Hence, we only need to prove that based on received $M$, with high
probability it is impossible for wiretapper to achieve distortion
less than $D_{E}$, if $R_{H}$ satisfies \eqref{eq:Thm1-2}.
\begin{thm}
\label{thm:Thm2} Let $\tau_{n}$ be any sequence that converges to
zero sub-exponentially fast (i.e., $\tau_{n}=2^{-o\left(n\right)}$).
If
\begin{equation}
{R_{H}\le\Gamma\left(R_{K},D_{E}\right)}\label{46)}
\end{equation}
then for $\forall\epsilon>0$,
\begin{align}
 & \mathop{\lim}\limits _{n\to\infty}\mathbb{P}_{{\mathcal{C}}_{S}}\Big[\mathop{\limsup}\limits _{l\to\infty}\mathbb{E}_{M^{l}}\nonumber \\
 & \mathop{\max_{\begin{array}{c}
\left(l,R'_{H}\right)Hcodes:\\
R'_{H}\leq R_{H}-\epsilon
\end{array}}}\mathbb{P}\left[d\left(S^{nl},Z^{nl}\right)\le D_{E}\right]>\tau_{n}\Big]=0
\end{align}
where $\Gamma\left(\cdotp\right)$ is defined in \eqref{eq:Gamma},
and $M^{l}=\left(M_{1},M_{2},\cdots,M_{l}\right)$.
\end{thm}
The proof is given in Appendix \ref{sec:Proof-of-Theorem2}.

Then by Theorem~\ref{thm:Thm2} we can establish the following corollary.
\begin{cor}
\label{Cor} If $R_{H}$ satisfies \eqref{eq:Thm1-2}, then for $\forall\epsilon>0$,
\begin{align}
 & \mathop{\lim}\limits _{n\to\infty}\mathbb{E}_{{\mathcal{C}}_{S}}\Big[\mathop{\limsup}\limits _{l\to\infty}\mathbb{E}_{M^{l}}\nonumber \\
 & \mathop{\max_{\begin{array}{c}
\left(l,R'_{H}\right)Hcodes:\\
R'_{H}\leq R_{H}-\epsilon
\end{array}}}\mathbb{P}\left[d\left(S^{nl},Z^{nl}\right)\le D_{E}\right]\Big]=0\label{eq:Conc}
\end{align}

\end{cor}

The Corollary \ref{Cor} implies \eqref{eq:Thm1-2} of Theorem~\ref{thm:Thm1}.
Hence the proof is completed.

\section{Causal Encryption System}

Here we consider a more general secrecy system, ``causal encryption
system''. Consider the secrecy communication system in Fig. \ref{fig:Shannon},
where the key is delivered at rate $R_{K}$, and the sender can not
only use the current key but also all past keys to code the current
block. We assume that the wiretapper has no decryption-delay constraint,
hence it can do decryption after receving all transmitted blocks (assume
total number is $L$). Besides, since the sender could adopt such
an adaptive encryption scheme (amount of key could be different for
different blocks), we assume the henchman and the wiretapper could
also arrange any $l$ transmitted blocks to form a superblock.

Let the sender only use the current key, then this kind of general
secrecy system is degenerated into the system described in Section
\ref{sec:Problem-Formulation}. By using the same compression and
encryption scheme in Section \ref{sec:Achievability-of-Theorem}, the sender
and the receiver could achieve the same secrecy performance. Hence
for causal encryption system, the achievablity of Theorem \ref{thm:Thm1}
still holds. In addition, for the first $\left(1-\lambda\right)L$
blocks, the henchman can use rate-$R_{K}$ henchman code to help the
wiretapper produce a lossless reconstruction (with probability 1),
and for the last $\lambda L$ blocks, the henchman can use rate-$R\left(D_{E}\right)$
henchman code to help the wiretapper produce a reconstruction of $S^{n}$
within distortion $D_{E}$ (with probability 1). Hence, the average
henchman rate is $\left(1-\lambda\right)R_{K}+\lambda R\left(D_{E}\right)$,
and the average distortion is $\lambda D_{E}$. This means that the
converse part of Theorem \ref{thm:Thm1} still holds as well. Hence
the achievable $\left(R,R_{K},R_{H},D_{E}\right)$ for this case is
also given by Theorem \ref{thm:Thm1}.

\section{Conclusion}
In this paper, we introduce a new measure on secrecy, \emph{security rate}, which is established based
on rate-distortion theory, and denotes the minimum (infimum) of the additional rate needed to reconstruct
the source within target distortion level with any positive probability
for wiretapper.
We study it in Shannon cipher system with lossless communication, and characterize
the admissible region of secret key rate, coding rate of the source,
wiretapper distortion, and security rate (distortion-based equivocation).
The single block version of henchman problem has been studied in \cite{Schieler14} and \cite{Schieler14-2}. In addition, by applying time-sharing decryption strategy, in the superblock version, the henchman and the wiretapper
are able to achieve any distortion-rate pair on or above the convex
hull of $\left(D,R_H \right)$ of the single block version. Moreover, we prove that such time-sharing decryption strategy is optimal. In addition, since the security rate equals the distortion-based equivocation, $\mathop{\min}\limits _{p\left(v^{n}|s^{n},m\right):Ed\left(S^{n},V^{n}\right)\le D_{E}}\frac{1}{n}I\left(S^{n};V^{n}|M\right)$,
and the equivocation is a special case of the distortion-based equivocation
(with Hamming distortion measure and $D_{E}=0$), this gives an answer
for the meaning of the maximum equivocation.

\newpage
\appendices{ }

\section{\label{sec:Properties-of-Distortion}Property of Distortion Set}
\begin{prop}
\label{prop:Property-of-Distortion}(Property of Distortion Set):
For any distortion measure $d(s^{n},z^{n})$, if both $|\mathcal{S}|$
and $|\mathcal{Z}|$ are finite, then the cardinality of $\mathcal{D}^{n}\triangleq\left\{ d(s^{n},z^{n}):s^{n}\in\mathcal{S}^{n},z^{n}\in\mathcal{Z}^{n}\right\} $
is at most polynomial in $n$.\end{prop}
\begin{IEEEproof}
Observe that the term $d(s^{n},z^{n})$ only depends on the joint
type of $(s^{n},z^{n})$, and moreover the number of joint types is
at most $\left(n+1\right)^{|\mathcal{S}||\mathcal{Z}|}$ \cite{Csiszar81},
which is polynomial in $n$.
\end{IEEEproof}

\section{\label{sec:Proof-of-Theorem2}Proof of Theorem \ref{thm:Thm2}}
\begin{IEEEproof}
We first define four envents
\begin{align}
 & \Acal_{1}\triangleq\left\{ S^{n}\in\Tcal_{\delta}^{n},S^{n}\in{\mathcal{C}}_{S}\right\} ,\\
 & \Acal_{2}\triangleq\left\{ \begin{array}{l}
\mathop{\max}\limits _{j_{p}\in\left[2^{n\left(R-R_{K}\right)}\right]}\mathop{\max}\limits _{z^{n}\in Z^{n}}\eta_{{\mathcal{C}}_{S},D}\left(z^{n},j_{p}\right)\\
\leq2^{n\left(\left[R_{K}-R(D)\right]^{+}+\epsilon\right)},\forall D\in\mathcal{D}^{n}
\end{array}\right\} ,\\
 & \Acal_{3}\triangleq\left\{ \mathop{\min}\limits _{j_{p}\in\left[2^{n\left(R-R_{K}\right)}\right]}\gamma_{{\mathcal{C}}_{S}}\left(j_{p}\right)\geq\left(1-\epsilon\right)2^{nR_{K}}\right\} ,\\
 & \Acal_{4}\triangleq\left\{ \begin{array}{l}
\mathop{\max}\limits _{s^{n}\in\Tcal_{\delta}^{n}}\phi_{{\mathcal{C}}_{S}}\left(s^{n}\right)\leq2^{n\left(R-H\left(S\right)+\epsilon\right)},\\
\mathop{\min}\limits _{s^{n}\in\Tcal_{\delta}^{n}}\phi_{{\mathcal{C}}_{S}}\left(s^{n}\right)\geq2^{n\left(R-H\left(S\right)-\epsilon\right)}
\end{array}\right\} ,
\end{align}
where
\begin{align}
\eta_{{\mathcal{C}}_{S},D}\left(z^{n},j_{p}\right) & \triangleq\sum_{j_{s}=1}^{2^{nR_{K}}}1\biggl\{ d\left(S^{n}\left(j_{p},j_{s}\right),z^{n}\right)\leq D,S^{n}\left(j_{p},j_{s}\right)\in\Tcal_{\delta}^{n}\biggr\}\\
\mathcal{D}^{n} & \triangleq\left\{ d(s^{n},z^{n}):s^{n}\in\mathcal{S}^{n},z^{n}\in\mathcal{Z}^{n}\right\} \\
\gamma_{{\mathcal{C}}_{S}}\left(j_{p}\right) & \triangleq\sum_{j_{s}=1}^{2^{nR_{K}}}1\left\{ S^{n}\left(j_{p},j_{s}\right)\in\Tcal_{\delta}^{n}\right\} \\
\phi_{{\mathcal{C}}_{S}}\left(s^{n}\right) & \triangleq\sum_{j\in\left[2^{nR}\right]}1\left\{ S^{n}\left(j\right)=s^{n}\right\}
\end{align}
Obviously the events $\Acal_{2},\Acal_{3},\Acal_{4}$ are only about
${\mathcal{C}}_{S}$, while the event $\Acal_{1}$ is about $S^{n}$
and ${\mathcal{C}}_{S}$. The $\delta$-typical set is defined according
to the notion of strong typicality, see \cite{Gamal}:

\begin{equation}
\Tcal_{\delta}^{n}(S)\triangleq\{s^{n}\in\Scal^{n}:\left|T_{s^{n}}-P_{S}\right|<\delta P_{S}\},
\end{equation}
where $T_{s^{n}}$ denotes the empirical distribution of $s^{n}$ (i.e., the type of $s^{n}$).

Next we will prove that any codebook ${\mathcal{C}}_{S}$ that makes
the events $\Acal_{2},\Acal_{3},\Acal_{4}$ hold will also make the
wiretapper decrypt the source with vanish probability. First, by method
of types, we can prove the Lemmas \ref{lem:A2}-\ref{lem:A4}.

\begin{lem}
\noindent \label{lem:A2}For any $\epsilon>0$ and small enough\textup{
$\delta$, }$\Pbb_{{\mathcal{C}}_{S}}\left[\overline{\Acal_{2}}\right]\to0$,
as $n\to\infty$.
\end{lem}

\begin{lem}
\noindent \label{lem:A3} For any $\epsilon>0$ and small enough\textup{
$\delta$,} $\Pbb_{{\mathcal{C}}_{S}}\left[\overline{\Acal_{3}}\right]\to0$,
as $n\to\infty$.
\end{lem}

\begin{lem}
\noindent \label{lem:A4} For any $\epsilon>0$ and small enough\textup{
$\delta$,} $\Pbb_{{\mathcal{C}}_{S}}\left[\overline{\Acal_{4}}\right]\to0$,
as $n\to\infty$. \end{lem}

The proofs of Lemmas \ref{lem:A2}, \ref{lem:A3} and \ref{lem:A4}
are given in Appendixes \ref{sec:Proof-of-LemmaA2}, \ref{sec:Proof-of-LemmaA3}
and \ref{sec:Proof-of-LemmaA4}, respectively.

In addition, denote

\begin{equation}
Q_{i}\triangleq\begin{cases}
1, & \textrm{if Node A encodes \ensuremath{i}th subblock successfully}\\
0\text{,} & \textrm{otherwise}
\end{cases}\label{typical-1}
\end{equation}
Observe that Node A encodes $i$th subblock successfully, if and only
if $\Acal_{1}$ happens; otherwise, randomly choose $J$. Therefore,

\begin{equation}
\Pbb\left(Q_{i}|{\mathcal{C}}_{S},\Acal\right)=\begin{cases}
\Pbb{}_{S^{n}}\left[\Acal_{1}|{\mathcal{C}}_{S},\Acal\right], & \textrm{if }Q_{i}=1\\
\Pbb{}_{S^{n}}\left[\overline{\Acal_{1}}|{\mathcal{C}}_{S},\Acal\right]\text{,} & \textrm{otherwise}
\end{cases}\label{typical-1-1}
\end{equation}
where $\Acal\triangleq\Acal_{2}\Acal_{3}\Acal_{4}$.
\begin{lem}
\noindent \label{lem:A5-1}For any $\delta>0$, if $R\geq H\left(S\right)+\epsilon$,
then $\Pbb{}_{S^{n}}\left[\overline{\Acal_{1}}|{\mathcal{C}}_{S},\Acal\right]\to0$,
as $n\to\infty$.
\end{lem}
The proof of Lemma \ref{lem:A5-1} is given in Appendix \ref{sec:Proof-of-LemmaA5-1}.

In the following, we denote $\epsilon_{n}\triangleq\Pbb{}_{S^{n}}\left[\overline{\Acal_{1}}|{\mathcal{C}}_{S},\Acal\right]$.
Observe that $Q^{l}$ is a sequence of i.i.d. random variables. Define

\begin{equation}
{\Acal_{5}\triangleq\left\{ Q^{l}\in\Tcal_{\delta}^{l}(Q)\right\} ,}\label{eq:-3}
\end{equation}
then by the property of typicality we have the following lemma.
\begin{lem}
\noindent \label{lem:A5}For any $\delta>0$, $\Pbb{}_{Q^{l}}\left[\overline{\Acal_{5}}|{\mathcal{C}}_{S},\Acal\right]\to0$,
as $l\to\infty$.
\end{lem}
We add $Q^{l}$ to the side information at both henchman and wiretapper.
Obviously, this will not decrease the decryption performance of henchman
and wiretapper. Therefore, to prove Theorem \ref{thm:Thm2}, we only
need to show
\begin{align*}
 & \mathop{\lim}\limits _{n\to\infty}\mathbb{P}_{{\mathcal{C}}_{S}}\Big[\mathop{\limsup}\limits _{l\to\infty}\mathbb{E}_{M^{l}Q^{l}}\\
 & \max_{Hcode\in\mathcal{H}_{l,R_{H}-\epsilon}}\Pbb[d\left(S^{nl},Z^{nl}\right)\leq D_{E}]>\tau_{n}\Big]=0
\end{align*}
i.e.,

\begin{align*}
 & \mathop{\lim}\limits _{n\to\infty}\mathbb{P}_{{\mathcal{C}}_{S}}\Biggl[\mathop{\limsup}\limits _{l\to\infty}\mathbb{E}_{M^{l}Q^{l}}\biggl[\\
 & \max_{Hcode\in\mathcal{H}_{l,R_{H}-\epsilon}}\mathbb{P}\left[d\left(S^{nl},Z^{nl}\right)\le D_{E}|M^{l}Q^{l}{\mathcal{C}}_{S}\right]|{\mathcal{C}}_{S}\biggr]>\tau_{n}\Biggr]=0
\end{align*}
where Hcodes denote the henchman codes with side information $M^{l}Q^{l}$
at both henchman and wiretapper.

First restrict the ${\mathcal{C}}_{S}$ to satisfy $\Acal$ by utilizing
Lemmas \ref{lem:A2}-\ref{lem:A4}.
\begin{align}
 & \mathbb{P}_{{\mathcal{C}}_{S}}\Big[\mathop{\limsup}\limits _{l\to\infty}\mathbb{E}_{M^{l}Q^{l}}\nonumber \\
 & \max_{Hcode\in\mathcal{H}_{l,R_{H}-\epsilon}}\Pbb[d\left(S^{nl},Z^{nl}\right)\leq D_{E}]>\tau_{n}\Big]\nonumber \\
\leq & \mathbb{P}_{{\mathcal{C}}_{S}}\Big[\mathop{\limsup}\limits _{l\to\infty}\mathbb{E}_{M^{l}Q^{l}}\nonumber \\
 & \max_{Hcode\in\mathcal{H}_{l,R_{H}-\epsilon}}\Pbb[d\left(S^{nl},Z^{nl}\right)\leq D_{E}]>\tau_{n}|\Acal\Big]+\mathbb{P}_{{\mathcal{C}}_{S}}[\overline{\Acal}]\nonumber \\
\leq & \mathbb{P}_{{\mathcal{C}}_{S}}\Big[\mathop{\limsup}\limits _{l\to\infty}\mathbb{E}_{M^{l}Q^{l}}\nonumber \\
 & \max_{Hcode\in\mathcal{H}_{l,R_{H}-\epsilon}}\Pbb[d\left(S^{nl},Z^{nl}\right)\leq D_{E}]>\tau_{n}|\Acal\Big]+\epsilon_{n}\label{eq:boundP}
\end{align}
where $\mathcal{H}_{l,R_{H}-\epsilon}\triangleq\left\{ \left(l,R'_{H}\right)Hcode:\, R'_{H}\leq R_{H}-\epsilon\right\} $.

We also restrict the $Q^{l}$ to satisfy $\Acal_{5}$ by utilizing
Lemma \ref{lem:A5}.

\begin{align*}
 & \mathop{\limsup}\limits _{l\to\infty}\mathbb{E}_{M^{l}Q^{l}}\biggl[\max_{Hcode\in\mathcal{H}_{l,R_{H}-\epsilon}}\\
 & \mathbb{P}\left[d\left(S^{nl},Z^{nl}\right)\le D_{E}|M^{l}Q^{l}\Acal{\mathcal{C}}_{S}\right]|\Acal{\mathcal{C}}_{S}\biggr]\\
\leq & \mathop{\limsup}\limits _{l\to\infty}\mathbb{E}_{M^{l}Q^{l}}\biggl[\max_{Hcode\in\mathcal{H}_{l,R_{H}-\epsilon}}\\
 & \mathbb{P}\left[d\left(S^{nl},Z^{nl}\right)\le D_{E}|M^{l}Q^{l}\Acal_{5}\Acal{\mathcal{C}}_{S}\right]|\Acal_{5}\Acal{\mathcal{C}}_{S}\biggr]\\
 & +\mathop{\limsup}\limits _{l\to\infty}\Pbb{}_{Q^{l}}\left[\overline{\Acal_{5}}|\Acal{\mathcal{C}}_{S}\right]\\
= & \mathop{\limsup}\limits _{l\to\infty}\mathbb{E}_{M^{l}Q^{l}}\biggl[\max_{Hcode\in\mathcal{H}_{l,R_{H}-\epsilon}}\\
 & \mathbb{P}\left[d\left(S^{nl},Z^{nl}\right)\le D_{E}|M^{l}Q^{l}\Acal_{5}\Acal{\mathcal{C}}_{S}\right]|\Acal_{5}\Acal{\mathcal{C}}_{S}\biggr]
\end{align*}

We can consider all possible sequences $z^{nl}$ given $(\Ccal_{S},M^{l}Q^{l})$ as a codebook. Since the wiretapper only receives $R_H$ rate message from the henchman, possible sequences $z^{nl}$ given $(\Ccal_{S},M^{l}Q^{l})$ produced by it is not more than $2^{nlR_H}$. Then a Hcode could be seen as the combination of a codebook (with size $2^{nlR_H}$) of $z^{n}$ sequences and an encoder that is designed based on that codebook.
Hence we can write

\begin{align}
 & \max_{Hcode\in\mathcal{H}_{l,R_{H}-\epsilon}}\Pbb[d(S^{nl},Z^{nl})\leq D_{E}|M^{l}Q^{l}\Acal_{5}\Acal{\mathcal{C}}_{S}]\nonumber \\
 & =\max_{c_{\zsf}(\Ccal_{S},M^{l}Q^{l})}\Pbb\Big[\min_{z^{nl}\in{c_{\zsf}(\Ccal_{S},M^{l}Q^{l})}}d(S^{nl},z^{nl})\leq D_{E}|M^{l}Q^{l}\Acal_{5}\Acal{\mathcal{C}}_{S}\Big],\label{eq:boundMax}
\end{align}
where the notation ${c_{\zsf}(\Ccal_{S},M^{l}Q^{l})}$ emphasizes that $c_{\zsf}$
is a function of the random codebook $\Ccal_{S}$ and public messages $(\Ccal_{S},M^{l}Q^{l})$.
Then by using a union bound, we can write the right-hand side of \eqref{eq:boundMax}
as

\begin{align}
 & \Pbb\Big[\min_{z^{nl}\in{c_{\zsf}(\Ccal_{S},M^{l}Q^{l})}}d(S^{nl},z^{nl})\leq D_{E}|M^{l}Q^{l}\Acal_{5}\Acal{\mathcal{C}}_{S}\Big]\nonumber \\
\stackrel{(a)}{\leq} & \sum_{z^{nl}\in{c_{\zsf}(\Ccal_{S},M^{l}Q^{l})}}\Pbb\Big[d(S^{nl},z^{nl})\leq D_{E}|M^{l}Q^{l}\Acal_{5}\Acal{\mathcal{C}}_{S}\Big]\\
\leq & 2^{nlR_{H}}\max_{z^{nl}\in{c_{\zsf}(\Ccal_{S},M^{l}Q^{l})}}\Pbb\Big[d(S^{nl},z^{nl})\leq D_{E}|M^{l}Q^{l}\Acal_{5}\Acal{\mathcal{C}}_{S}\Big]\\
\leq & 2^{nlR_{H}}\max_{z^{nl}\in\Zcal^{nl}}\Pbb\Big[d(S^{nl},z^{nl})\leq D_{E}|M^{l}Q^{l}\Acal_{5}\Acal{\mathcal{C}}_{S}\Big]\\
\overset{\left(b\right)}{\leq} & 2^{nlR_{H}}\max_{z^{nl}\in\Zcal^{nl}}\Pbb\Big[\sum_{i=1}^{t}d(S_{i}^{n}(J_{i}),z_{i}^{n})\leq lD_{E}|M^{l}Q^{l}\Acal_{5}\Acal{\mathcal{C}}_{S}\Big]\\
= & 2^{nlR_{H}}\max_{z^{nl}\in\Zcal^{nl}}\Pbb\Big[\sum_{i=1}^{t}d(S_{i}^{n}(J_{i}),z_{i}^{n})\leq lD_{E}|Q^{t}=1,M^{t},\Acal,{\mathcal{C}}_{S}\Big]\\
= & 2^{nlR_{H}}\max_{z^{nt}\in\Zcal^{nt}}\sum_{j_{1},\cdots,j_{t}}\prod_{i=1}^{t}\Pbb\left[J_{i}=j_{i}|Q_{i}=1,M_{i}=m_{i},\Acal,{\mathcal{C}}_{S}\right]\nonumber \\
 & 1\left\{ \sum_{i=1}^{t}d(S_{i}^{n}(J_{i}),z_{i}^{n})\leq lD_{E}\right\} \label{eq:boundMin}
\end{align}
where step (a) is a union bound, and step (b) follows from that 1)
by definition of typicality, the number of ''1'' in $Q^{l}$ is
at least $l\left(1-\delta\right)\left(1-\epsilon_{n}\right)$, and
$t\triangleq l\left(1-\delta\right)\left(1-\epsilon_{n}\right)$,
2) without loss of generalization we assume $Q_{i}=1,1\leq i\leq t$.
Now we derive the probability $\Pbb\left[J_{i}=j_{i}|Q_{i}=1,M_{i}=m_{i}\right]$.
For different $i$, $\left(J_{i},M_{i},Q_{i}\right)$ is i.i.d., hence
for simplicity, we use $\Pbb\left[J=j|Q=1,M=m\right]$ to denote $\Pbb\left[J_{i}=j_{i}|Q_{i}=1,M_{i}=m_{i}\right]$.
Moreover, for simplicity, we also omit condition $\Acal,{\mathcal{C}}_{S}$
for each probability expression. First we have

\begin{align}
 & {\mathbb{P}\left[S^{n}=s^{n},J=j,Q=1,M=m\right]}\nonumber \\
= & {\mathbb{P}\left[S^{n}=s^{n},J=j,M=m,S^{n}\in\Tcal_{\delta}^{n},S^{n}\in{\mathcal{C}}_{S}\right]}\\
= & \mathbb{P}\left[S^{n}=s^{n},S^{n}\in\Tcal_{\delta}^{n},S^{n}\in{\mathcal{C}}_{S}\right]\nonumber \\
 & \times\mathbb{P}\left[J=j|S^{n}=s^{n},S^{n}\in\Tcal_{\delta}^{n},S^{n}\in{\mathcal{C}}_{S}\right]\mathbb{P}\left[M=m|J=j\right].\label{eq:-4}
\end{align}
For first term of \ref{eq:-4}, we have
\begin{align}
 & \mathbb{P}\left[S^{n}=s^{n},S^{n}\in\Tcal_{\delta}^{n},S^{n}\in{\mathcal{C}}_{S}\right]\nonumber \\
= & \mathbb{P}\left[S^{n}=s^{n}\right]1\left\{ s^{n}\in\Tcal_{\delta}^{n},s^{n}\in{\mathcal{C}}_{S}\right\}
\end{align}
and

\begin{equation}
\begin{array}{l}
2^{-n\left(H\left(S\right)+\epsilon\right)}\end{array}\le\Pbb\left[S^{n}=s^{n}\right]\le2^{-n\left(H\left(S\right)-\epsilon\right)},ifs^{n}\in\Tcal_{\delta}^{n}\label{eq:-5}
\end{equation}
where \eqref{eq:-5} is the property of typicality \cite{Gamal}.
For other terms of \ref{eq:-4}, we have
\begin{align}
 & \mathbb{P}\left[J=j|S^{n}=s^{n},S^{n}\in\Tcal_{\delta}^{n},S^{n}\in{\mathcal{C}}_{S}\right]\nonumber \\
\geq & 2^{-n\left(R-H\left(S\right)+\epsilon\right)}1\left\{ S^{n}\left(j\right)=s^{n}\right\} \label{eq:-6}
\end{align}

\begin{align}
 & \mathbb{P}\left[J=j|S^{n}=s^{n},S^{n}\in\Tcal_{\delta}^{n},S^{n}\in{\mathcal{C}}_{S}\right]\nonumber \\
\leq & 2^{-n\left(R-H\left(S\right)-\epsilon\right)}1\left\{ S^{n}\left(j\right)=s^{n}\right\} \label{eq:-8}
\end{align}
and

\begin{equation}
\mathbb{P}\left[M=m|J=j\right]=2^{-nR_{K}}1\left\{ m_{p}=j_{p}\right\}
\end{equation}
where \eqref{eq:-6} and \eqref{eq:-8} follow from the encoding process
and the condition $\mathcal{C}_{S}$ satisfying $\Acal_{4}$.

Therefore,

\begin{align}
 & {\mathbb{P}\left[J=j|M=m,Q=1\right]}\nonumber \\
= & \frac{{\sum_{s^{n}\in\mathcal{S}^{n}}\mathbb{P}\left[S^{n}=s^{n},J=j,Q=1,M=m\right]}}{\sum_{j\in\left[2^{nR}\right]}\sum_{s^{n}\in\mathcal{S}^{n}}\mathbb{P}\left[S^{n}=s^{n},J=j,Q=1,M=m\right]}\\
\leq & \frac{\sum_{s^{n}\in\mathcal{S}^{n}}2^{-n\left(H\left(S\right)-\epsilon\right)}1\left\{ s^{n}\in\Tcal_{\delta}^{n},s^{n}\in{\mathcal{C}}_{S}\right\} ...}{\sum_{j\in\left[2^{nR}\right]}\sum_{s^{n}\in\mathcal{S}^{n}}2^{-n\left(H\left(S\right)+\epsilon\right)}1\left\{ s^{n}\in\Tcal_{\delta}^{n},s^{n}\in{\mathcal{C}}_{S}\right\} ...}\nonumber \\
 & \frac{2^{-n\left(R-H\left(S\right)-\epsilon\right)}1\left\{ S^{n}\left(j\right)=s^{n}\right\} 2^{-nR_{K}}1\left\{ m_{p}=j_{p}\right\} }{2^{-n\left(R-H\left(S\right)+\epsilon\right)}1\left\{ S^{n}\left(j\right)=s^{n}\right\} 2^{-nR_{K}}1\left\{ m_{p}=j_{p}\right\} }\\
\leq & \frac{2^{n4\epsilon}1\left\{ S^{n}\left(j\right)\in\Tcal_{\delta}^{n},S^{n}\left(j\right)\in{\mathcal{C}}_{S}\right\} 1\left\{ m_{p}=j_{p}\right\} }{\sum_{j\in\left[2^{nR}\right]}1\left\{ S^{n}\left(j\right)\in\Tcal_{\delta}^{n},S^{n}\left(j\right)\in{\mathcal{C}}_{S}\right\} 1\left\{ m_{p}=j_{p}\right\} }\\
= & \frac{2^{n4\epsilon}1\left\{ S^{n}\left(j\right)\in\Tcal_{\delta}^{n}\right\} 1\left\{ m_{p}=j_{p}\right\} }{\sum_{j\in\left[2^{nR}\right]}1\left\{ S^{n}\left(j\right)\in\Tcal_{\delta}^{n}\right\} 1\left\{ m_{p}=j_{p}\right\} }\\
\leq & \frac{2^{n4\epsilon}1\left\{ S^{n}\left(j\right)\in\Tcal_{\delta}^{n}\right\} 1\left\{ m_{p}=j_{p}\right\} }{\sum_{j_{s}\in\left[2^{nR_{K}}\right]}1\left\{ S^{n}\left(m_{p},j_{s}\right)\in\Tcal_{\delta}^{n}\right\} }\\
\leq & \frac{2^{n4\epsilon}1\left\{ S^{n}\left(j\right)\in\Tcal_{\delta}^{n}\right\} 1\left\{ m_{p}=j_{p}\right\} }{\left(1-\epsilon\right)2^{nR_{K}}}\label{eq:-7}
\end{align}
where \eqref{eq:-7} follows from the condition $\mathcal{C}_{S}$
satisfying $\Acal_{3}$.

Combining \eqref{eq:boundMin} and \eqref{eq:-7}, we have

\begin{align}
 & \Pbb\Big[\min_{z^{nl}\in{c_{\zsf}(\Ccal_{S},M^{l}Q^{l})}}d(S^{nl},z^{nl})\leq D_{E}|M^{l}Q^{l}\Acal_{5}\Acal{\mathcal{C}}_{S}\Big]\nonumber \\
\leq & \mu\max_{z^{nt}\in\Zcal^{nt}}\sum_{j_{s,1},\cdots,j_{s,t}}\prod_{i=1}^{t}1\left\{ S^{n}\left(m_{p,i},j_{s,i}\right)\in\Tcal_{\delta}^{n}\right\} \nonumber \\
 & 1\{\sum_{i=1}^{t}d(S^{n}(m_{p,i},j_{s,i}),z_{i}^{n})\leq lD_{E}\}\\
\leq & \mu\max_{z^{nt}\in\Zcal^{nt}}\sum_{j_{s,1}=1}^{2^{nR_{K}}}\sum_{j_{s,2}=1}^{2^{nR_{K}}}\cdots\sum_{j_{s,t}=1}^{2^{nR_{K}}}\sum_{D_{1},D_{2},\cdots,D_{t}:\sum_{i=1}^{t}D_{i}\leq lD_{E}}\nonumber \\
 & \prod_{i=1}^{t}1\{d(S^{n}(m_{p,i},j_{s,i}),z_{i}^{n})=D_{i},S^{n}(m_{p,i},j_{s,i})\in\Tcal_{\delta}^{n}\}\\
\leq & \mu\sum_{D_{1},D_{2},\cdots,D_{t}:\sum_{i=1}^{t}D_{i}\leq lD_{E}}\eta_{1}\eta_{2}\cdots\eta_{t}\\
\stackrel{(a)}{=} & \mu2^{O\left(l\log n\right)}\max_{D_{1},D_{2},\cdots,D_{t}:\sum_{i=1}^{t}D_{i}\leq lD_{E}}\eta_{1}\eta_{2}\cdots\eta_{t}\label{eq:boundMin2}
\end{align}
where $\mu=\frac{1}{\left(1-\epsilon\right)^{t}}2^{nlR_{H}-nt\left(R_{K}-4\epsilon\right)}$,
\begin{align}
\eta_{i}\triangleq & \mathop{\max}\limits _{m_{p,i}\in\left[2^{n\left(R-R_{K}\right)}\right]}\max_{z_{i}^{n}\in\Zcal^{n}}\label{eq:Eta}\\
 & \sum_{j_{s,i}=1}^{2^{nR_{K}}}1\{d(S^{n}(m_{p,i},j_{s,i}),z_{i}^{n})\leq D_{i},S^{n}(m_{p,i},j_{s,i})\in\Tcal_{\delta}^{n}\}
\end{align}
and step (a) follows from that the number of the distortion incurred
by $n$ length block is at most polynomial in $n$, see Appendix \ref{sec:Properties-of-Distortion}.

Combining \eqref{eq:boundP}, \eqref{eq:boundMax}, and \eqref{eq:boundMin2},
we have \eqref{eq:PEta} (at top of the next page), where $\mu_{2}=\tau_{n}\left(1-\epsilon\right)^{t}2^{nt\left(R_{K}-4\epsilon\right)-nlR_{H}-O\left(l\log n\right)}$.
\newcounter{mytempeqncnt} 
\begin{figure*}[!t]
\begin{align}
 & \mathbb{P}_{{\mathcal{C}}_{S}}\Big[\mathop{\limsup}\limits _{l\to\infty}\mathbb{E}_{M^{l}Q^{l}}\max_{Hcode\in\mathcal{H}_{l,R_{H}-\epsilon}}\Pbb[d(S^{nl},Z^{nl})\leq D_{E}]>\tau_{n}\Big]\nonumber \\
 & \leq\mathbb{P}_{{\mathcal{C}}_{S}}\Biggl[\mathop{\limsup}\limits _{l\to\infty}\mathbb{E}_{M^{l}Q^{l}}\biggl[\max_{c_{\zsf}(\Ccal_{S},M^{l}Q^{l})}\Pbb\Big[\min_{z^{nl}\in{c_{\zsf}(\Ccal_{S},M^{l}Q^{l})}}d(S^{nl},z^{nl})\leq D_{E}|M^{l}Q^{l}\Acal_{5}\Acal{\mathcal{C}}_{S}\Big]|\Acal_{5}\Acal{\mathcal{C}}_{S}\biggr]>\tau_{n}|\Acal\Biggr]+\epsilon_{n}\nonumber \\
 & \leq\mathbb{P}_{{\mathcal{C}}_{S}}\Big[\mathop{\limsup}\limits _{l\to\infty}\mathbb{E}_{M^{l}Q^{l}}\biggl[\max_{D_{1},D_{2},\cdots,D_{t}:\sum_{i=1}^{t}D_{i}\leq lD_{E}}\eta_{1}\eta_{2}\cdots\eta_{t}|\Acal_{5}\Acal{\mathcal{C}}_{S}\biggr]>\mu_{2}|\Acal\Big]+\epsilon_{n}\nonumber \\
 & =\mathbb{P}_{{\mathcal{C}}_{S}}\Big[\mathop{\limsup}\limits _{l\to\infty}\max_{D_{1},D_{2},\cdots,D_{t}:\sum_{i=1}^{t}D_{i}\leq lD_{E}}\eta_{1}\eta_{2}\cdots\eta_{t}>\mu_{2}|\Acal\Big]+\epsilon_{n}\label{eq:PEta}
\end{align}
\addtocounter{equation}{1} 

\end{figure*}

Therefore, we only need to prove when ${\mathcal{C}}_{S}=c$ make
$\Acal_{2}$ hold, then there exists a large enough $n$ (not dependent
on $l$) that makes the following hold for any $l$.

\begin{equation}
\max_{D_{1},D_{2},\cdots,D_{t}:\sum_{i=1}^{t}D_{i}\leq lD_{E}}\eta_{1}\eta_{2}\cdots\eta_{t}\leq\mu_{2}\label{eq:PEta2}
\end{equation}
Assume $\Acal_{2}$ holds, then we have

\begin{align}
 & \max_{D_{1},D_{2},\cdots,D_{t}:\sum_{i=1}^{t}D_{i}\leq lD_{E}}\eta_{1}\eta_{2}\cdots\eta_{t}\nonumber \\
\leq & \max_{D_{1},D_{2},\cdots,D_{t}:\sum_{i=1}^{t}D_{i}\leq lD_{E}}\prod_{i=1}^{t}2^{n\left(\left[R_{K}-R(D_{i})\right]^{+}+\epsilon\right)}
\end{align}
Hence we only need show
\begin{equation}
\max_{D_{1},D_{2},\cdots,D_{t}:\sum_{i=1}^{t}D_{i}\leq lD_{E}}\prod_{i=1}^{t}2^{n\left(\left[R_{K}-R(D_{i})\right]^{+}+\epsilon\right)}\leq\mu_{2}\label{eq:PEta2-1}
\end{equation}
Assume there are $b$ of $D_{i}$ such that $R_{K}>R(D_{i})$ in $t$
subblocks, and $\lambda\triangleq\frac{b}{t}$. Then we have
\begin{align}
 & \max_{D_{1},D_{2},\cdots,D_{t}:\sum_{i=1}^{t}D_{i}\leq lD_{E}}\prod_{i=1}^{t}2^{n\left(\left[R_{K}-R(D_{i})\right]^{+}+\epsilon\right)}\nonumber \\
 & \leq\max_{\lambda,D_{1},D_{2},\cdots,D_{\lambda t}:\sum_{i=1}^{\lambda t}D_{i}\leq lD_{E}}2^{tn\epsilon}2^{\lambda tnR_{K}}2^{-n\sum_{i=1}^{\lambda t}R(D_{i})}\\
 & \leq\max_{\lambda,D:\lambda tD\leq lD_{E}}2^{tn\epsilon}2^{\lambda tn\left(R_{K}-R(D)\right)}\\
 & =2^{tn\epsilon+\max_{\lambda,D:\lambda tD\leq lD_{E}}\left[\lambda tn\left(R_{K}-R(D)\right)\right]}
\end{align}
For sufficiently $n$, all subexponential terms in $\mu_{2}$ can
be omitted, hence we only need show
\[
2^{tn\epsilon+\max_{\lambda,D:\lambda tD\leq lD_{E}}\left[\lambda tn\left(R_{K}-R(D)\right)\right]}\leq2^{nt\left(R_{K}-4\epsilon\right)-nlR_{H}}
\]
i.e.,
\begin{align}
R_{H} & \leq\frac{1}{l}\left[t\left(R_{K}-4\epsilon\right)-t\epsilon-\max_{\lambda,D:\lambda tD\leq lD_{E}}\left[\lambda t\left(R_{K}-R(D)\right)\right]\right]\nonumber \\
 & =\frac{t}{l}\left[-5\epsilon+\min_{\lambda,D:\lambda D\leq\frac{t}{l}D_{E}}\left[\left(1-\lambda\right)R_{K}+\lambda R(D)\right]\right]\label{eq:-10}
\end{align}

For small enough $\delta$, $\epsilon$ and large enough $n$, we
have

\[
\frac{t}{l}=\left(1-\delta\right)\left(1-\epsilon_{n}\right)\rightarrow1
\]
Then \eqref{eq:-10} becomes

\begin{align*}
R_{H} & \leq\min_{\lambda,D:\lambda D\leq D_{E}}\left[\left(1-\lambda\right)R_{K}+\lambda R(D)\right]
\end{align*}
This is just the condition of Theorem \eqref{thm:Thm2}. Hence the
achievability of Theorem~\ref{thm:Thm2} holds.
\end{IEEEproof}

\section{\label{sec:Proof-of-LemmaA2}Proof of Lemma \ref{lem:A2}}
\begin{IEEEproof}
\noindent According to definition of $\Acal_{2}$,
\begin{align}
 & {\Pbb_{{\mathcal{C}}_{S}}\left[\overline{\Acal_{2}}\right]}\\
= & {\Pbb_{{\mathcal{C}}_{S}}\left[\begin{array}{l}
\mathop{\max}\limits _{j_{p}\in\left[2^{n\left(R-R_{K}\right)}\right]}\mathop{\max}\limits _{z^{n}\in Z^{n}}\eta_{{\mathcal{C}}_{S},D}\left(z^{n},j_{p}\right)\\
>2^{n\left(\left[R_{K}-R(D)\right]^{+}+\epsilon\right)},\exists D\in\mathcal{D}^{n}
\end{array}\right]}\\
\le & \left|\mathcal{Z}^{n}\right|2^{n\left(R-R_{K}\right)+O\left(\log n\right)}\mathop{\max}\limits _{j_{p}\in\left[2^{n\left(R-R_{K}\right)}\right]}\mathop{\max}\limits _{D\in\mathcal{D}^{n}}\mathop{\max}\limits _{z^{n}\in Z^{n}}\nonumber \\
 & {\Pbb_{{\mathcal{C}}_{S}}\left[\eta_{{\mathcal{C}}_{S},D}\left(z^{n},j_{p}\right)>2^{n\left(\left[R_{K}-R(D)\right]^{+}+\epsilon\right)}\right]}\label{eq:A2_UnionBound}
\end{align}

\noindent where \eqref{eq:A2_UnionBound} follows from a union bound
and the fact $\left|\mathcal{D}^{n}\right|$ is polynomial in $n$,
see Appendix \ref{sec:Properties-of-Distortion}. Define $\xi_{j_{p},j_{s},z^{n}}\triangleq1\biggl\{ d\left(S^{n}\left(j_{p},j_{s}\right),z^{n}\right)\leq D,S^{n}\left(j_{p},j_{s}\right)\in\Tcal_{\delta}^{n}\biggr\}$,
then $\eta_{z^{n},{\mathcal{C}}_{S}\left(j_{p}\right)}=\sum_{j_{s}=1}^{2^{nR_{K}}}\xi_{j_{p},j_{s},z^{n}}$.
Given $j_{p}$ and $z^{n}$, $\xi_{j_{p},j_{s},z^{n}},j_{s}\in\left[2^{nR_{K}}\right]$
are i.i.d. random variables, with mean

\noindent
\begin{equation}
\begin{array}{l}
\begin{array}{c}
\Ebb_{{\mathcal{C}}_{S}}\xi_{j_{p},j_{s},z^{n}}\end{array}=\Pbb\left\{ d\left(S^{n},z^{n}\right)\leq D_{E},S^{n}\in\Tcal_{\delta}^{n}\right\} \end{array}
\end{equation}

If we can show that the probability in \eqref{eq:A2_UnionBound} decays
doubly exponentially fast with $n$, then the proof will be complete.
To that end, we first introduce the following lemmas.
\begin{lem}
\label{lem:typebound}\cite{Schieler14}, \cite{Schieler14-2} If
$S^{n}$ is i.i.d. according to $P_{S}$, then for any $z^{n}$,
\begin{equation}
\Pbb[d(S^{n},z^{n})\leq D,S^{n}\in\Tcal_{\delta}^{n}]\leq2^{-n(R(D)-o(1))},\label{eq:rd-1-1}
\end{equation}
where $R(D)$is the point-to-point rate-distortion function for $P_{S}$
, and $o(1)$ is a term that vanishes as $\delta\rightarrow0$ and
$n\rightarrow\infty$ .
\end{lem}

\begin{lem}
\label{lem:chernoff} \cite{Schieler14}, \cite{Schieler14-2} If
$X^{m}$ is a sequence of i.i.d. $\text{Bern}(p)$ random variables,
then
\begin{equation}
\Pbb\Big[\sum_{i=1}^{m}X_{i}>k\Big]\leq\left(\frac{emp}{k}\right)^{k}.
\end{equation}
\end{lem}

\noindent From Lemma~\ref{lem:typebound}, we see that

\begin{equation}
\begin{array}{l}
\begin{array}{c}
\Ebb_{{\mathcal{C}}_{S}}\xi_{j_{p},j_{s},z^{n}}\end{array}\leq2^{{-n(R(D)-o(1))}}\end{array}
\end{equation}

\noindent Using the bound on $\Ebb[\xi_{j_{p},j_{s},z^{n}}]$, we
can apply Lemma~\ref{lem:chernoff} to the probability in \eqref{eq:A2_UnionBound}
by identifying
\begin{equation}
\begin{array}{l}
m=2^{{nR_{K}}}\\
p\leq2^{{-n(R(D)-o(1))}}\\
k=2^{n\left(\left[R_{K}-R(D)\right]^{+}+\epsilon\right)}.
\end{array}\label{eq:-2-1-1}
\end{equation}

\noindent This gives
\begin{equation}
\Pbb\Big[\sum_{j=1}^{2^{nR_{K}}}\xi_{j_{p},j_{s},z^{n}}>2^{n\left(\left[R_{K}-R(D)\right]^{+}+\epsilon\right)}\Big]\leq2^{-n\alpha2^{n\beta}},\label{doubleexp}
\end{equation}
where
\begin{equation}
\begin{array}{l}
\alpha=\left[R_{K}-R(D)\right]^{+}-\left(R_{K}-R(D)\right)+\epsilon-o(1)\\
\beta=\left[R_{K}-R(D)\right]^{+}+\epsilon.
\end{array}\label{eq:-1-1-1-1}
\end{equation}
For any fixed $\epsilon$ and large enough $n$, both $\alpha$ and
$\beta$ are positive and bounded away from zero, and \eqref{doubleexp}
vanishes doubly exponentially fast. Consequently, the expression in
\eqref{eq:A2_UnionBound} vanishes, completing the proof of Lemma~\ref{lem:A2}.
\end{IEEEproof}

\section{\label{sec:Proof-of-LemmaA3}Proof of Lemma \ref{lem:A3}}
\begin{IEEEproof}
\noindent According to definition of $\Acal_{3}$,
\begin{align}
 & \Pbb_{{\mathcal{C}}_{S}}\left[\overline{\Acal_{3}}\right]\\
 & =\Pbb_{{\mathcal{C}}_{S}}\Biggl[\mathop{\min}\limits _{j_{p}\in\left[2^{n\left(R-R_{K}\right)}\right]}\gamma_{{\mathcal{C}}_{S}\left(j_{p}\right)}<\left(1-\epsilon\right)2^{nR_{K}}\Biggr]\\
 & \le2^{n\left(R-R_{K}\right)}\mathop{\max}\limits _{j_{p}\in\left[2^{n\left(R-R_{K}\right)}\right]}\Pbb_{{\mathcal{C}}_{S}}\left[\gamma_{{\mathcal{C}}_{S}\left(j_{p}\right)}<\left(1-\epsilon\right)2^{nR_{K}}\right]\label{eq:A3_UnionBound}
\end{align}
Define $\zeta_{j_{p},j_{s}}\triangleq1\left\{ S^{n}\left(j_{p},j_{s}\right)\in\Tcal_{\delta}^{n}\right\} $,
then $\gamma_{{\mathcal{C}}_{S}\left(j_{p}\right)}=\sum_{j_{s}=1}^{2^{nR_{K}}}\zeta_{j_{p},j_{s}}$.
Given $j_{p}$ , $\zeta_{j_{p},j_{s}},j_{s}\in\left[2^{nR_{K}}\right]$
are i.i.d. random variables, with mean

\noindent
\begin{align}
 & \Ebb_{{\mathcal{C}}_{S}}\zeta_{j_{p},j_{s}}\\
 & =\Pbb\left\{ S^{n}\left(j\right)\in\Tcal_{\delta}^{n}\right\} \\
 & \geq1-\epsilon_{2}\label{eq:A3Mean-2}
\end{align}

\noindent where \eqref{eq:A3Mean-2} holds for any $\epsilon_{2}>0$
and sufficiently large $n$, and it follows from the typicality property.
If we can show that the probability in \eqref{eq:A3_UnionBound} decays
doubly exponentially fast with $n$, then the proof will be complete.
To that end, we first introduce the following lemma on Chernoff bound.
\begin{lem}
\label{lem:chernoff2}\cite{Mitzenmacher} If $X^{m}$ is a sequence
of i.i.d. $\text{Bern}(p)$ random variables, then
\begin{equation}
\Pbb\Big[\sum_{i=1}^{m}X_{i}\leq(1-\delta)mp\Big]\leq e^{-\frac{\delta^{2}mp}{2}},0\le\delta\le1.
\end{equation}

\end{lem}
By indentifying that
\begin{align}
m & =2^{nR_{K}}\label{121)-2}\\
p & \ge1-\epsilon_{2}\\
\delta & \geq1-\frac{1-\epsilon}{1-\epsilon_{2}}
\end{align}
and applying Lemma \ref{lem:chernoff2}, we have
\begin{equation}
\Pbb_{{\mathcal{C}}_{S}}\left\{ \gamma_{{\mathcal{C}}_{S}\left(j_{p}\right)}<\left(1-\epsilon\right)2^{nR_{K}}\right\} \le e^{-\frac{\delta^{2}mp}{2}}\le e^{-\frac{\delta^{2}m\left(1-\epsilon_{2}\right)}{2}}\label{eq:A3_DoubleExp-2}
\end{equation}

For fixed $\epsilon$, if $\delta$ is small enough and $n$ is large
enough, then $\epsilon_{2}<\epsilon$ and $\epsilon_{2}<1$, which
means that $\delta>0$ and $1-\epsilon_{2}>0$. Hence \eqref{eq:A3_DoubleExp-2}
vanishes doubly exponentially fast. This completes the proof of Lemma
\ref{lem:A3}.
\end{IEEEproof}

\section{\label{sec:Proof-of-LemmaA4}Proof of Lemma \ref{lem:A4}}
\begin{IEEEproof}
\noindent According to definition of $\Acal_{4}$,

\noindent
\begin{align}
 & \Pbb_{{\mathcal{C}}_{S}}\left[\overline{\Acal_{4}}\right]\\
 & =\Pbb_{{\mathcal{C}}_{S}}\left[\begin{array}{l}
\mathop{\max}\limits _{s^{n}\in\Tcal_{\delta}^{n}}\phi_{{\mathcal{C}}_{S}}\left(s^{n}\right)>2^{n\left(R-H\left(S\right)+\epsilon\right)},\\
or\,\mathop{\min}\limits _{s^{n}\in\Tcal_{\delta}^{n}}\phi_{{\mathcal{C}}_{S}}\left(s^{n}\right)<2^{n\left(R-H\left(S\right)-\epsilon\right)}
\end{array}\right]\\
 & \le\Pbb_{{\mathcal{C}}_{S}}\left[\begin{array}{l}
\mathop{\max}\limits _{s^{n}\in\Tcal_{\delta}^{n}}\phi_{{\mathcal{C}}_{S}}\left(s^{n}\right)>2^{n\left(R-H\left(S\right)+\epsilon\right)}\end{array}\right]\nonumber \\
 & +\Pbb_{{\mathcal{C}}_{S}}\left[\begin{array}{l}
\mathop{\min}\limits _{s^{n}\in\Tcal_{\delta}^{n}}\phi_{{\mathcal{C}}_{S}}\left(s^{n}\right)<2^{n\left(R-H\left(S\right)-\epsilon\right)}\end{array}\right]\label{eq:-9}
\end{align}

\noindent In the following, we prove that both terms of \eqref{eq:-9}
vanish as $n\rightarrow\infty$.

\noindent
\begin{align}
 & \Pbb_{{\mathcal{C}}_{S}}\left[\begin{array}{l}
\mathop{\max}\limits _{s^{n}\in\Tcal_{\delta}^{n}}\phi_{{\mathcal{C}}_{S}}\left(s^{n}\right)>2^{n\left(R-H\left(S\right)+\epsilon\right)}\end{array}\right]\nonumber \\
\le & \left|\Tcal_{\delta}^{n}\right|\mathop{\max}\limits _{s^{n}\in\Tcal_{\delta}^{n}}\Pbb_{{\mathcal{C}}_{S}}\left[\phi_{{\mathcal{C}}_{S}}\left(s^{n}\right)>2^{n\left(R-H\left(S\right)+\epsilon\right)}\right]\label{eq:A4_UnionBound1}
\end{align}
Define $\zeta_{j,s^{n}}\triangleq1\left\{ S^{n}\left(j\right)=s^{n}\right\} $,
then $\phi_{{\mathcal{C}}_{S}}\left(s^{n}\right)=\sum_{j\in\left[2^{nR}\right]}\zeta_{j,s^{n}}$.
Given $s^{n}\in\Tcal_{\delta}^{n}$ , $\zeta_{j,s^{n}},j\in\left[2^{nR}\right]$
are i.i.d. random variables, with mean

\noindent
\begin{align}
 & \Ebb_{{\mathcal{C}}_{S}}\zeta_{j,s^{n}}\\
 & =\Pbb\left\{ S^{n}\left(j\right)=s^{n}\right\} \\
 & \leq2^{-n\left(H\left(S\right)-\epsilon_{2}\left(\delta\right)\right)}\label{eq:A4Mean}
\end{align}

\noindent where $\epsilon_{2}\left(\delta\right)$ tends to zero as
$\delta\rightarrow0$, and \eqref{eq:A4Mean} follows from the typicality
property \cite{Gamal}. If we can show that the probability in \eqref{eq:A4_UnionBound1}
decays doubly exponentially fast with $n$, then the proof will be
complete.

By indentifying that
\begin{align}
m & =2^{nR}\label{121)}\\
p & \leq2^{-n\left(H\left(S\right)-\epsilon_{2}\left(\delta\right)\right)}\\
k & =2^{n\left(R-H\left(S\right)+\epsilon\right)}
\end{align}
and applying Lemma \ref{lem:chernoff}, we have
\begin{equation}
\Pbb_{{\mathcal{C}}_{S}}\left[\phi_{{\mathcal{C}}_{S}}\left(s^{n}\right)>2^{n\left(R-H\left(S\right)+\epsilon\right)}\right]\le2^{-n\alpha2^{n\beta}},\label{eq:A4_DoubleExp}
\end{equation}

\noindent where
\begin{equation}
\begin{array}{l}
\alpha=\epsilon-\epsilon_{2}\left(\delta\right)\\
\beta=R-H\left(S\right)+\epsilon.
\end{array}\label{eq:-1-1-1-1-2}
\end{equation}

For fixed $\epsilon$, if $n$ is large enough and $\delta$ is small
enough, then $\epsilon_{2}\left(\delta\right)<\epsilon$, which means
that $\alpha>0$ and $\beta>0$. Hence \eqref{eq:A4_DoubleExp} vanishes
doubly exponentially fast. This means that \eqref{eq:A4_UnionBound1}
holds.

In the same way, by utilizing Lemma \ref{lem:chernoff2}, we can prove
that for small enough $\delta$, as $n\rightarrow\infty$,

\begin{equation}
\Pbb_{{\mathcal{C}}_{S}}\left[\begin{array}{l}
\mathop{\min}\limits _{s^{n}\in\Tcal_{\delta}^{n}}\phi_{{\mathcal{C}}_{S}}\left(s^{n}\right)<2^{n\left(R-H\left(S\right)-\epsilon\right)}\end{array}\right]\rightarrow0
\end{equation}
Therefore, this completes the proof of Lemma \ref{lem:A4}.
\end{IEEEproof}

\section{\label{sec:Proof-of-LemmaA5-1}Proof of Lemma \ref{lem:A5-1}}
\begin{IEEEproof}
\noindent According to definition of $\Acal_{1}$, we have

\noindent
\begin{align}
 & \Pbb{}_{S^{n}}\left[\Acal_{1}|{\mathcal{C}}_{S},\Acal\right]\\
 & =\Pbb{}_{S^{n}}\left[S^{n}\in\Tcal_{\delta}^{n},S^{n}\in{\mathcal{C}}_{S}|{\mathcal{C}}_{S},\Acal\right]\\
 & =\Pbb{}_{S^{n}}\left[S^{n}\in\Tcal_{\delta}^{n}|{\mathcal{C}}_{S},\Acal\right]\Pbb{}_{S^{n}}\left[S^{n}\in{\mathcal{C}}_{S}|S^{n}\in\Tcal_{\delta}^{n},{\mathcal{C}}_{S},\Acal\right]\label{eq:-9-2}\\
 & =\Pbb{}_{S^{n}}\left[S^{n}\in\Tcal_{\delta}^{n}|{\mathcal{C}}_{S},\Acal\right]\\
 & =\Pbb{}_{S^{n}}\left[S^{n}\in\Tcal_{\delta}^{n}\right]\label{eq:-12}
\end{align}

\noindent where \eqref{eq:-12} follows from that if $R\geq H\left(S\right)+\epsilon$,
$S^{n}\in\Tcal_{\delta}^{n}$ and ${\mathcal{C}}_{S}$ satisfying
$\Acal_{4}$, then $S^{n}\in{\mathcal{C}}_{S}$. In addition, by law
of large numbers, we have $\Pbb{}_{S^{n}}\left[S^{n}\in\Tcal_{\delta}^{n}\right]\rightarrow1$
as $n\rightarrow\infty$ \cite{Gamal}.\end{IEEEproof}

\end{document}